# Implicit reporting standards in bibliometric research: what can reviewers' comments tell us about reporting completeness?

Dimity Stephen[1], Alexander Schniedermann[1], Andrey Lovakov[1], Marion Schmidt[1], Matteo Ottaviani[1], Nikita Sorgatz[1,2], Roberto Cruz Romero[1], Torger Möller[1], Valeria Aman[1], and Stephan Stahlschmidt[1,3]

DS: *stephen@dzhw.eu*, ORCID: 0000-0002-7787-6081
AS: *schniedermann@dzhw.eu*, ORCID: 0000-0003-2132-7419
AL: *lovakov@dzhw.eu*, ORCID: 0000-0001-8644-9236
MS: *schmidt@dzhw.eu*, ORCID: 0000-0001-6970-3714
MO: *ottaviani@dzhw.eu*, ORCID: 0009-0004-2339-082X
NS: *nikita.sorgatz@hu-berlin.de,* ORCID: 0000-0003-3878-6267
RCR: *cruzromero@dzhw.eu*, ORCID: 0000-0001-5740-7037
TM: *moeller@dzhw.eu,* ORCID: 0009-0003-9753-8295
VA: *aman@dzhw.eu,* ORCID: 0000-0001-8370-780X
SS: *stahlschmidt@dzhw.eu*, ORCID: 0000-0003-3390-8632

[1] German Centre for Higher Education Research and Science Studies, Berlin, Germany
[2] Robert K. Merton Center for Science Studies, Humboldt-Universität zu Berlin, Berlin, Germany
[3] Unit of Computational Humanities and Social Sciences (U-CHASS), EC3 Research Group, University of Granada, Granada, Spain

The recent surge in bibliometric studies published has been accompanied by increasing diversity in the completeness of reporting these studies' details, affecting reliability, reproducibility, and robustness. Our study systematises the reporting of bibliometric research using open peer reviews. We examined 182 peer reviews of 85 bibliometric studies published in library and information science (LIS) journals and conference proceedings, and non-LIS journals. We extracted 968 reviewer comments and inductively classified them into 11 broad thematic categories and 68 sub-categories, determining that reviewers largely focus on the completeness and clarity of reporting data, methods, and results. We subsequently derived 49 recommendations for the details authors should report and compared them with the GLOBAL, PRIBA, and BIBLIO reporting guidelines to identify (dis)similarities in content. Our recommendations addressed 60-80% of the guidelines' items, while the guidelines covered 45-65% of our recommendations. Our recommendations provided greater range and specificity, but did not incorporate the functions of guidelines beyond addressing academic content. We argue that peer reviews provide valuable information for the development of future guidelines. Further, our recommendations can be read as the implicit community standards for reporting bibliometric studies and could be used by authors to aid complete and accurate reporting of their manuscripts.

**Key words:** reporting guidelines; peer review; bibliometrics; library and information science

**Declarations**

**Data availability statement**
This study utilised bibliometric studies published open access and with open peer review. All data used in the analyses and the results are available here: https://docs.google.com/spreadsheets/d/13Lib6ipGXxF7xxemHIFog4IExLH9kjDE/edit?usp=sharing&ouid=103789543211253180375&rtpof=true&sd=true . A final version will be made available on Zenodo following peer review.

**Competing interests**
The authors have no competing interests. However, we wish to note that three authors (DS, AS, SS) participated in the development of the GLOBAL reporting guidelines.

**Funding information**
No external funding was received for this study.

**Author contributions**
Conceptualization: DS, AS, AL, MS, MO, NS, RCR, TM, VA, SS. Data curation: DS, AS, AL, MS, MO, NS, RCR, TM, SS, VA. Data analysis: DS. Writing – original draft: DS. Writing – review and editing: AL, AS, DS, MS, NS, SS, TM, VA.

# 1. Introduction

Publication output in the field of bibliometrics is growing at an unchecked rate. Larivière (2012) and Jonkers and Derrick (2012) detected a sudden spurt in bibliometric studies in 2003 and growth has only accelerated since then: the number of publications increased 12-fold from around 800 in 2000-2004 to over 10,000 by 2015-2019 (González-Alcaide, 2021). Notably, the share of these studies published in library and information science (LIS) journals – the field historically central to bibliometrics – has steadily decreased over time from around 70% in the 1980s and 1990s to 40% in 2010 (Larivière, 2012), then to around 25% in 2019 (González-Alcaide, 2021). Life science and medicine and social science journals now publish the lion's share of bibliometric studies, with around a third each (González-Alcaide, 2021).

This rapid growth in bibliometric studies may be attributed to several diverse factors. For instance, the prominence of bibliometrics in international, national, and institutional research evaluation and management activities (Cabezas-Clavijo et al., 2023; González-Alcaide, 2021) has raised its profile amongst scholars in all fields. Further, the increasing availability of data sources and analytical software has made bibliometrics accessible to everyone (Cabezas-Clavijo et al., 2023; Boyack, Klavans & Smith, 2022). Viewed cynically, these advances have opened the field to "academic opportunists", who perceive bibliometric analyses as a quick and easy approach to boosting their publication output (González-Alcaide, 2021). Viewed positively, the self-monitoring capacity has been empowered substantially in all research fields. From either perspective, the prominence and accessibility of bibliometrics have thus generated a wave of interest in the field across disciplines.

While the widespread uptake of our methods should be celebrated as an acknowledgment of our field's relevance and potential to contribute broadly to academia (Klavans & Boyack, 2022), if unchecked, it may also negatively impact the quality, rigour, and development of our field. For instance, our central theories and principles are unlikely to be known to researchers dropping by from other fields to borrow methods and data. Consequently, the bibliometric corpus may be inundated with studies that make minimal contributions to the field or misuse methods and indicators (Jonkers & Derrick, 2012; González-Alcaide, 2021). Individually, such studies are unlikely to have a notable impact on the field. However, in large numbers, they can collectively produce misleading effects, which damages both the theoretical advancement of our field and its reputation among academics and policy-makers (Boyack et al., 2022).

Well-documented data and methods are central to the reliability, reproducibility, and robustness of bibliometric studies (Boyack et al., 2022). Evidence of issues in the reporting of bibliometric studies remains currently rather anecdotal. However, a small number of studies that empirically examined reporting quality have found wide variation in the reporting of study characteristics, with – for instance – good reporting of search terms, but poor reporting of database characteristics (Koo & Lin, 2023); that substantial numbers of studies lacked the sufficient detail necessary for replicating their findings (Boyack et al., 2022); and that under-reporting of methodological details was widespread in studies both within and outside the LIS field (Cabezas-Clavijo et al., 2023). These findings suggest that the broad community of scholars using bibliometrics could benefit from the guidance in the responsible and effective use of bibliometric data and methods that has long been called for (e.g., Glänzel & Schoepflin, 1994; Glänzel, 1996; González-Alcaide, 2021).

First steps toward providing this guidance are being made with the development of three independent reporting guidelines that aim to assist authors with adequately reporting bibliometric studies. The first of these guidelines to be published was the Preferred Reporting Items for Bibliometric Analysis (PRIBA) by Koo and Lin (2023). The PRIBA items were derived from the Preferred Reporting Items for Systematic reviews and Meta-Analyses (PRISMA) 2020 guidelines and adapted by the authors for bibliometric analyses (Koo & Lin, 2023). The two other guidelines – the Guidance List for repOrting Bibliometric AnaLyses (GLOBAL) by Ng et al. (2023) and the Guideline for Reporting Bibliometric Reviews of the Biomedical Literature (BIBLIO) by Montazeri et al. (2023) – were developed through similar processes: first a scoping review of existing reporting recommendations to identify potential items (Ng et al., 2025), and then a Delphi consensus process with individuals experienced with bibliometric analyses to finalise the guidelines' content (Ng et al., 2023; Montazeri et al., 2023).

Notably, each of these guidelines were developed using a deductive perspective based on recommendations in the existing literature and supplemented with expert experience. We do not criticise the methodological development of the three guidelines. Indeed, GLOBAL was developed in accordance with current best practice for guideline development in health research (Moher, Schulz, Simera, & Altman, 2010; Enhancing the QUAlity and Transparency Of health Research (EQUATOR) Network, n.d.; Ng, et al., 2023). Instead, we engage with the question of what guideline items might be generated using, conversely, an inductive perspective. In this study, we viewed peer reviewers' comments about specific bibliometric studies as reflections of the implicit community standards for the reporting of such studies and interpret them as reporting recommendations. Thus, contrasting the guideline content developed using these complementary methodologies might raise (dis)similiar perspectives of what is recommended to be reported. Further, examining content from an inductive perspective may highlight issues that are central to the reporting of bibliometric studies, but that are unable to be adequately captured in the guideline format. Establishing and maintaining this continuously evolving set of shared concepts via efforts such as developing reporting guidelines not only facilitates scholarly communication, laying the groundwork for progress, but also has the potential to shape education and training in bibliometric methods. As such, a comparison of the recommended reporting practices identified using differing and complementary methodologies is crucial to ensuring reporting guidelines are well-founded, and to understanding what recommendations are (necessarily) missing from reporting guidelines.

*1.1 Research aims and approach*

The aim of this study is to i) explore and systematise issues in the quality and completeness of reporting bibliometric research and ii) compare the reporting recommendations in existing guidelines developed deductively with recommendations derived based on an inductive approach. We thus address two questions. First, what aspects of reporting are mentioned by peer reviewers in their reviews of bibliometric studies, and with what frequency? Second, what is the extent and focus of (dis)agreement between recommendations derived via complementary methods?

Broadly, our approach is to qualitatively examine open peer reviews of bibliometric studies and identify aspects that reviewers raise as well- or poorly reported. For example, reviewers may ask for additional information regarding databases, sample sizes, search terms, filter criteria, of the indicators used, suggesting the provided details were insufficient for understanding or reproducing the study. Thus, our inductive and descriptive approach focuses on issues that have been identified by diverse academic peers in open peer review procedures at both central and peripheral bibliometric publishing outlets. This approach facilitates a quantitative, albeit non-systematic, inspection of the prevalence of particular reporting issues. More importantly, however, from these reviews we can develop implicit recommendations for authors reporting their bibliometric studies. We then qualitatively compare these inductively-derived recommendations with the three existing bibliometric reporting guidelines (GLOBAL, PRIBA, BIBLIO) to examine the (dis)similiarities in content, and identify aspects of reporting that peer reviewers identify as central to reporting bibliometric studies, and those that cannot be adequately captured in reporting guidelines.

## 2. Methods

The methodology used in this study consists of four key phases, which are detailed in the following sections and outlined in Figure 1. Briefly, we first identified three samples of bibliometric studies and retrieved the peer reviews for these documents. In the second phase, we examined these reviews to extract reviewers' critiques and commendations, and then qualitatively analysed these comments in a three-step process that inductively generated implicit reporting recommendations. We then undertook a quantitative analysis of the prevalence of reporting issues, and finally, compared the recommendations generated with the PRIBA, GLOBAL, and BIBLIO reporting guidelines to identify (dis)similarities.

*2.1 Identification and sampling of bibliometric studies*

We defined a bibliometric study as a study that used a bibliometric data source (e.g., Web of Science, Scopus) and one or more metadata fields (e.g., journal, discipline) to compare two or more entities or groups (e.g., authors, institutions, countries) to contribute knowledge to its field (e.g., one database covers more journals than another or one institution is more productive than another). We decided to use three samples of bibliometric studies: articles in LIS journals, articles in non-LIS journals, and submissions to a LIS conference. As noted, a substantial number of bibliometric studies now appear in non-LIS journals and these may be reviewed by peers with a different perspective of the details necessary to sufficiently report a bibliometric study than reviewers of articles in LIS journals, which potentially increases the diversity of aspects raised. Similarly, conference papers are usually shorter in length than articles and may contain fewer methodological details and results than articles, and so reviewers may highlight particularly important features when these are missing. This sample of both articles and conference submissions may thus capture a wide array of issues raised by reviewers, aligning with the interdisciplinarity present in bibliometrics.

For each sample, we first identified potential bibliometric studies in journals and conference proceedings with open peer review and then applied our definition of a bibliometric study. We required that the peer review process for the submission was open so that we could examine the reviewers' comments. The 27th International Conference on Science, Technology and Innovation Indicators (STI 2023) had an open peer review process, so we used these submissions to represent conference proceedings (sample A in Figure 1). To identify bibliometric studies in these submissions, we extracted the title, abstract and keywords (TAK) of each submission from the Orvium website using the rvest (Wickham, 2022a), tidyverse (Wickham, et al. 2019), jsonlite (Ooms, 2014), and stringr (Wickham, 2022b) R packages. We then narrowed the submissions to those that used a bibliometric data source by searching for any of the following (case insensitive) terms in the TAK: Web of Science, WoS, Scopus, Dimensions, Openalex, Open Alex, Pubmed, Crossref, SciELO, Wikidata, Overton, altmetric, bibliometric data, DOAJ. We then manually screened the full-texts of these submissions to assess whether they fulfilled the aforementioned criteria of using a metadata field and comparing groups to make a knowledge claim. Those that fulfilled these criteria were retained as bibliometric studies, and we downloaded their peer reviews from the Orvium website.

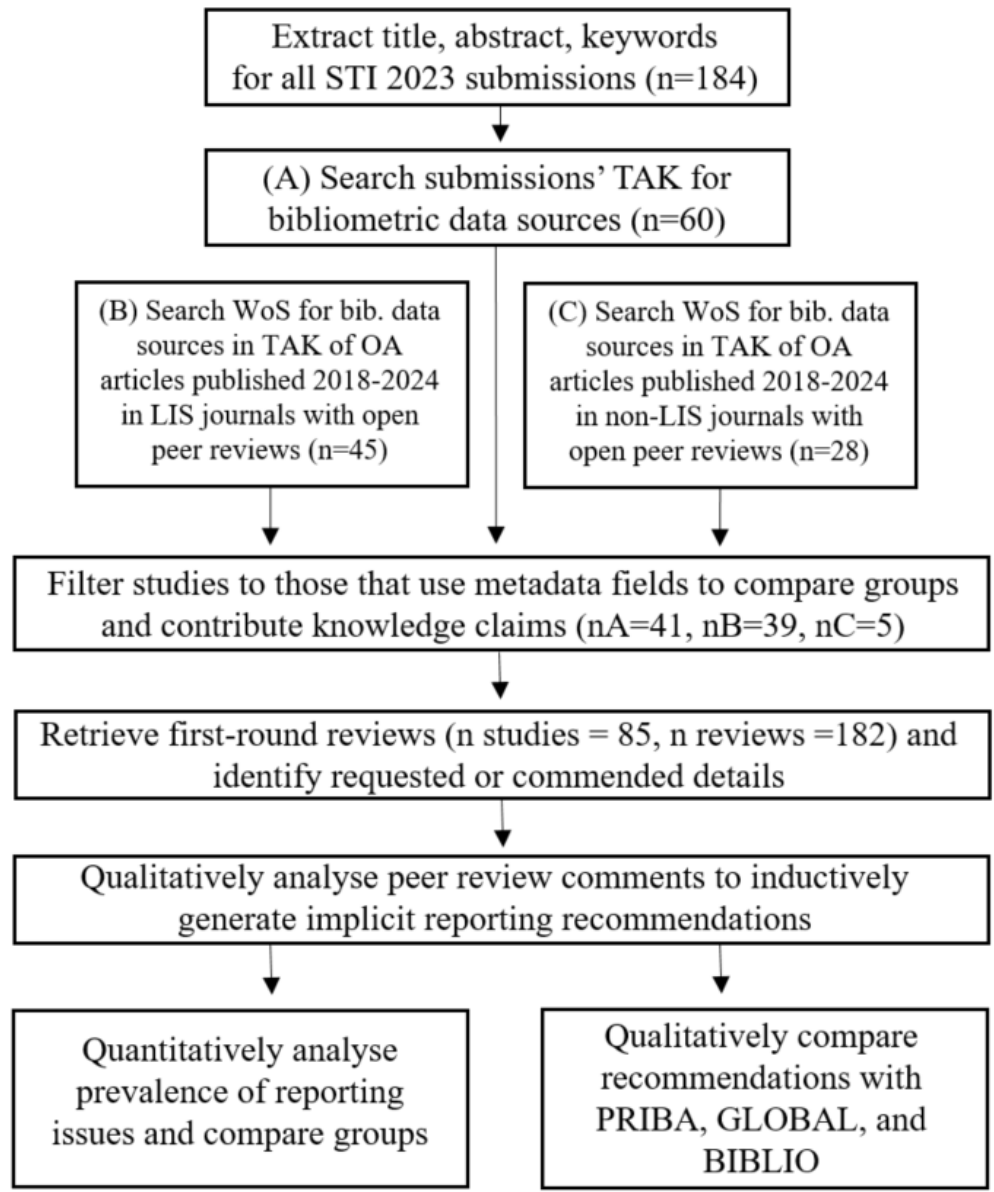

Figure 1: Flowchart of method. TAK refers to Title, Abstract, Keywords.

To identify NLIS and LIS articles with open peer reviews, we performed two searches of the online WoS database on 29 February 2024, including the Science Citation Index Expanded, the Social Sciences Citation Index, and the Arts & Humanities Citation Index. First, to identify LIS articles (sample B), we searched for any of the aforementioned bibliometrics data sources in the Topic (TAK) field. In addition, we restricted the publication years to 2018-2024, the WoS Subject Category to "Information & Library Science", the document type to article, and filtered the results to those articles that were open access (OA) and had open peer reviews available. We performed the same search to identify non-LIS articles (sample C), with the following changes: Category was not "Information & Library Science", the title did not include "Protocol", Dimensions and Pubmed were removed, and "scientometric" was added as a search term, as we observed authors to use bibliometric and scientometric interchangeably. "Protocol" was excluded to remove study protocols. Dimensions was removed because it is unlikely to refer to the database outside of LIS, and Pubmed was removed as its inclusion returned many out-of-scope systematic reviews. For both samples, we then manually screened the studies' full-texts to retain those that fulfilled our criteria as bibliometric studies. We then downloaded the first-round peer reviews for all bibliometric studies in samples B and C via the "Open Peer Reviews" link in WoS.

*2.2 Qualitative analysis of reviewer comments*

To prepare for our qualitative analysis of the peer reviewers' comments, we first extracted all relevant comments from the reviewers' reports for all three samples. This included all remarks about the theoretical framing or methodological detail of the study, but not, for instance, summaries of the study provided for the editor. The comments could be positive, such as praise for a clear or detailed description of the methodology; negative, such as critiquing the study's limitations; or neutral, such as suggestions for additional references. In this process, each team member examined and extracted comments from approximately 20 peer reviews.

We then undertook a three-stage iterative process of categorising comments into themes, each stage with a greater level of specificity, with the goal of ultimately developing reporting recommendations from the reviewers' comments. In the first stage, we categorised the comments into broad themes based on the overarching concept of the comment. Here, in an inductive process, we discussed the comments' focus and identified and allocated comments to a broad category (e.g., description of results, visualisations/tables, clarity and validity of concepts). Comments could be assigned to more than one category. In the second stage, we then repeated this process for the comments within each category and identified a set of more specific sub-categories. For instance, the reviewer comment "What is the unit for y-axis in Figure 7?" was first assigned to the broad category of Visualisations and Tables and then sub-categorised to (Un)clear presentation. Sub-categories were labelled neutrally as comments could be positive, neutral, or negative. Once classified, all comments in each sub-category were reviewed for internal consistency and reclassified to other or new sub-categories as required. In this way, we inductively classified all comments to both broad categories and more specific sub-categories based on the concept addressed in the comment. In the third stage, we translated the key aspects of each of the sub-categories into a reporting recommendation, phrased as "Authors should …" that sought to capture the core focus of the reviewers' comments. For instance, seven comments from reviewers that referred to undefined acronyms (e.g., "Clarify the acronym: JUOJS", "Use of the OA abbreviation is inconsistent, open access is spelled out even in the next mention after presenting the abbreviation", "Abbreviations [e.g., IR in page 11, MA in page 14] should be spelled out the full term at its first mention and indicated its abbreviation in parenthesis") were first categorised to the broad category "Clarity of presentation", then the sub-category "Explain acronym", and then translated to the reporting recommendation "Authors should define acronyms at the first mention, then apply them consistently throughout the paper", which captured the key points of defining acronyms, defining them at the first mention, and using them consistently thereafter. Finally, we reviewed the complete list of reporting recommendations for clarity of wording and specificity of the concept and split out compound items or merged items with similar content.

*2.3 Quantitative analysis of reviewer comments*

To quantify the frequency at which reviewers commented on the various aspects of reporting, we calculated the number and percentage of comments that pertained to each broad and sub-category of themes and the nature of these comments (positive, negative, or neutral). Further, to examine whether there were any differences between samples, we further disambiguated these data by document type (article from N/LIS journal or conference submission). Similarly, we also examined the number and percentage of documents that received positive or negative comments in each of the broad and sub-categories, and if any differences occurred by document type. The small number of NLIS articles available prevented a reliable analysis comparing LIS article, NLIS article, or LIS conference submissions to identify any differences between N/LIS peer reviews.

*2.4 Qualitative comparison of reporting recommendations with existing guidelines*

We compared the reporting recommendations developed in this study against the three existing guidelines for bibliometric studies. First, we provide further details about the focus, content, and development of these guidelines. The GLOBAL seeks to guide authors in reporting bibliometric studies that appear in journals, without a specific orientation toward a particular discipline or target group. GLOBAL was developed in two steps: a scoping literature review to identify items, and a two-round Delphi process to refine and finalise the items based on expert opinion. In the first step, the developers searched seven bibliographic databases, preprint servers, and grey literature using specific search terms to identify documents published until July 2023 that contained guidance on reporting bibliometric analyses. They screened the resulting 48,750 records and identified 23 such documents (including both PRIBA and BIBLIO), generating a list of 32 potential guideline items (Ng, et al., 2025). In the Delphi process, the developers first surveyed 145 self-selected participants about the necessity of reporting each item. Items that received 80% or more of the votes to include or exclude were treated accordingly. In a second round, a panel of 16 bibliometric experts reached a verdict on items that did not achieve consensus in the first round. The guideline thus consists of 29 items distributed over the abstract (1 item), introduction (4 items), methods (13 items), results (4 items), discussion (3 items), and other (4 items), which encompasses declarations, data statements, and references (Ng, et al., 2023). Notably, we examined the current version of the GLOBAL, which is undergoing pilot testing, and so some changes may exist between the version used here and the final version of GLOBAL when it is published.

The BIBLIO guidelines were developed similarly to GLOBAL. The developers first searched the PubMed, Scopus, WoS, and Cochrane Library databases for documents published until September 2023 that contained a combination of various keywords and MeSH terms in the titles. They retrieved and screened 13,720 documents, retaining 19 documents that contained guidance for reporting bibliometric studies, resulting in a preliminary list of 20 items. In a second step, these 20 items were presented to a panel of 11 experts, "including [a] bibliometrician, epidemiologist, clinician, librarian, statistician, journal editor, and a research fellow", who were surveyed to select which items, or suggest additional items, that should be included in the guidelines. In three further surveys, the developers asked the same panel to rate and re-rate the potential items on a 5-point scale, providing the results between iterations, until they reached consensus of 75% or more votes for an item's inclusion or exclusion (Montazeri, et al., 2023). The BIBLIO guidelines are thus comprised of 20 items that "should be described as a minimum requirements (sic) in reporting a bibliometric review [in the following document sections]: title (2 items), abstract (1 item), introduction/background (2 items), methods (7 items), results (4 items), discussion (4 items)" and are specifically oriented toward reporting bibliometric analyses of biomedical literature (Montazeri, et al., 2023).

Like BIBLIO, the PRIBA guidelines are also oriented toward providing guidance for reporting bibliometric studies in the health and medical sciences (Koo & Lin, 2023). Regarding the method of developing PRIBA, the developers stated that "[t]he individual items [in PRIBA] were modified and expanded from the ones used in the PRISMA 2020 Checklist for bibliometric studies" (Koo & Lin, 2023). Consequently, PRIBA largely mirrors the PRISMA guidelines, with its 25 items distributed across seven sections: title (2 items), abstract (5 items), introduction (2 items), methods (7 items), results (3 items), discussion (3 items), and other information (3 items), covering declarations and data availability.

To identify the themes common and unique to our list and the existing guidelines, we compared each item in our list against each of the guidelines to determine whether the same concept was also present in the guideline. To achieve the converse perspective, we then compared each item in the guidelines against our list to identify which concepts were (not) included in our list. Overarchingly, we decided that items that contained similar, if not identical, or partial aspects of the concept or intent were "matches", and that these differences would be explored in the qualitative analysis. For instance, we decided that "Provide an explicit statement of the objective(s)" in PRIBA was comparable to our item "Authors should clearly state the research question(s) that the study addresses" because both items address the same concept of identifying the aim of the study, although not in an identical manner. We calculated the number and percentage of items in our list present in the guidelines and vice versa as a simple quantitative summary. We also further qualitatively examined the items present and absent from the list and guidelines to identify the common or unique content.

## 3. Results

The total sample consisted of 182 reviews of 85 bibliometric studies: 8 reviews of 5 studies published in NLIS journals, 79 reviews of 39 studies published in LIS journals, and 95 reviews of 41 LIS conference papers. The low number of NLIS studies occurs as the open peer review restriction severely limited the sample to only six journals: Ecological Solutions and Evidence, Engineering Reports, Environmental Research Letters, Internet Technology Letters, Journal of Oral Rehabilitation, and Royal Society Open Science. Similarly, the LIS articles were all published in Quantitative Science Studies, as the only WoS-indexed LIS journal with open peer review, and all LIS conference papers were submitted to STI 2023.

On average, reviews of conference papers were 287 words in length (range = 31-1,091 words), which was – as could be expected – shorter than article reviews. Reviews of bibliometric studies in the LIS journals were notably longer (mean = 710 words, range = 76-2,605) than articles in NLIS journals (mean = 536 words, range = 35-2,062).

*3.1 Results of qualitative and quantitative analysis of reviewers' comments*

The initial extraction of comments from the reviews identified 1,030 relevant comments. Sixty-two comments were later deemed to be out of scope of the analysis and removed, leaving 968 comments in the analysis. Half of all comments were drawn from LIS article reviews (525 comments, 54.2% of all 968 comments), 392 (40.5%) from the STI reviews, and 51 (5.3%) from NLIS article reviews. The first process of categorising comments identified 11 broad themes: *Clarity and validity of concepts*; *Clarity of presentation*; *Description of data/methods*; *Description of results*; *Visualisations and tables*; *Limitations*; *Conclusions*; *Open Science/Reproducibility*; *Declarations*; *Links to literature/references*; and *Overall assessment*. Table 1 shows the total number and percentage of comments in these 11 categories and by comment type – negative (i.e. critical of the manuscript), neutral, or positive. As comments could be classified to more than one category, the total count of comments exceeds 968. Table S0 in the Supplementary Material (SM) contains the full dataset of document metadata, reviewer comments, and category allocations.

Table 1. The number and percentage of comments in the 11 broad categories and comment type, ordered by the total number of comments.

| Category | No. (%) comments | No. (%) negative | No. (%) neutral | No. (%) positive |
|---|---|---|---|---|
| Description of data / methods | 329 (29.1) | 287 (87.2) | 5 (1.5) | 37 (11.2) |
| Clarity of presentation | 139 (12.3) | 89 (64.3) | 0 (0.0) | 50 (35.7) |
| Visualisations and tables | 136 (12.0) | 118 (86.8) | 6 (4.4) | 12 (8.8) |
| Description of results | 131 (11.6) | 111 (84.7) | 11 (8.4) | 9 (6.9) |
| Overall assessment | 118 (10.5) | 51 (42.9) | 0 (0.0) | 67 (56.8) |
| Links to literature / references | 83 (7.4) | 69 (84.1) | 2 (2.4) | 12 (14.5) |
| Clarity and validity of concepts | 62 (5.5) | 57 (91.9) | 0 (0.0) | 5 (8.1) |
| Conclusions | 59 (5.2) | 54 (91.5) | 0 (0.0) | 5 (8.5) |
| Open science / reproducibility | 42 (3.7) | 20 (47.6) | 0 (0.0) | 22 (52.4) |
| Limitations | 29 (2.6) | 20 (69.0) | 2 (6.9) | 7 (24.1) |
| Declarations | 1 (0.1) | 1 (100.0) | 0 (0.0) | 0 (0.0) |
| Total | 1,129 (100) | 877 (77.7) | 26 (2.3) | 226 (20.0) |

Reviewers commented most often about the authors' description of the data and or methods used in the study, accounting for 29.1% (329) of comments. A substantial percentage of comments also pertained to the clarity of the information presented (139, 12.3%), the visualisations and tables used (136, 12.0%), and the description of results (131, 11.6%). Most of the comments in each category were critical, which aligns with the aim of peer review to identify potential issues and suggest improvements to the authors. However, notably, comments relating to open science and reproducibility (e.g., the provision of the data and or analytical scripts used in the study) and the overall assessment of the study (e.g., its contextualisation in the existing literature, appropriateness of its design to address the research question, and its originality, utility, and relevance) were more often positive than negative. This latter instance is likely influenced by the fact that all articles and 80.5% of the conference submissions (33 of 41) examined were eventually accepted for publication or presentation. Overall, the completeness and clarity of the methods and results of bibliometric studies thus constituted the core focus of reviewers' comments.

The second categorisation process identified 68 sub-categories of these 11 themes. The number and percentage of reviewers' comments pertaining to these sub-categories are shown in Table 2. In addition, Tables S1-S3 in the SM show the number of comments disambiguated by broad and sub-category, document type (article or conference paper), and comment type (negative, neutral, or positive), while Table S4 shows the number and percentage of *documents* that received comments in each of these sub-categories by comment type.

Table 2. The number and percentage of comments by sub-category of themes.

| Broad category<br>Sub-category | No. of comments | % of comments | % of broad category |
|---|---|---|---|
| **Clarity and validity of concepts** | **62** | **5.5** | **100.0** |
| Definition required | 35 | 3.1 | 56.5 |
| Disagreement with interpretation | 8 | 0.7 | 12.9 |
| Research design is (in)appropriate | 16 | 1.4 | 25.8 |
| Terminology | 3 | 0.3 | 4.8 |
| **Clarity of presentation** | **139** | **12.3** | **100.0** |
| (Un)clear research questions | 20 | 1.8 | 14.4 |
| (Un)clear text | 62 | 5.5 | 44.6 |
| Explain acronym | 7 | 0.6 | 5.0 |
| Inconsistency | 13 | 1.2 | 9.4 |
| Redundancy | 8 | 0.7 | 5.8 |
| Structure | 25 | 2.2 | 18.0 |
| Suggest alternate presentation | 4 | 0.4 | 2.9 |
| **Conclusions** | **59** | **5.2** | **100.0** |
| Conclusions are (un)founded | 31 | 2.7 | 52.5 |
| Framing issue | 5 | 0.4 | 8.5 |
| Framing issue due to research design issue | 4 | 0.4 | 6.8 |
| Generalizability & robustness | 13 | 1.2 | 22.0 |
| Request for qualification | 6 | 0.5 | 10.2 |
| **Declarations** | **1** | **0.1** | **100.0** |
| Conflict of interest | 1 | 01 | 100.0 |
| **Description of data/methods** | **329** | **29.1** | **100.0** |
| Communication/presentation | 2 | 0.2 | 0.6 |
| Details about data are (in)sufficient | 50 | 4.4 | 15.2 |
| Details about methods are (in)sufficient | 108 | 9.6 | 32.8 |
| Doubts about statistical soundness | 3 | 0.3 | 0.9 |
| Ex/inclusion criteria | 11 | 1.0 | 3.3 |
| Incorrect information | 5 | 0.4 | 1.5 |
| Justification of data | 16 | 1.4 | 4.9 |
| Justification of methods | 51 | 4.5 | 15.5 |
| Research design is (in)appropriate | 34 | 3.0 | 10.3 |
| Robustness | 13 | 1.2 | 4.0 |
| Underlying data | 36 | 3.2 | 10.9 |
| **Description of results** | **131** | **11.6** | **100.0** |
| (In)complete interpretation | 36 | 3.2 | 27.5 |
| (In)complete results | 14 | 1.2 | 10.7 |
| Communication/presentation | 4 | 0.4 | 3.1 |
| Doubts about statistical soundness | 14 | 1.2 | 10.7 |
| Extend/further analysis | 35 | 3.1 | 26.7 |
| Incorrect information | 1 | 0.1 | 0.8 |
| Research design is (in)appropriate | 13 | 1.2 | 9.9 |
| Sample size | 9 | 0.8 | 6.9 |
| Statistical reporting/communication | 5 | 0.4 | 3.8 |
| **Limitations** | **29** | **2.6** | **100.0** |
| Limitations were (not) discussed | 22 | 1.9 | 75.9 |
| Research design is (in)appropriate | 1 | 0.1 | 3.4 |
| Underlying data | 6 | 0.5 | 20.7 |
| **Links to literature / references** | **83** | **7.4** | **100.0** |
| Inappropriate link to previous literature | 6 | 0.5 | 7.2 |
| Literature review is (in)sufficient | 23 | 2.0 | 27.7 |
| Missing reference(s) | 11 | 1.0 | 13.3 |
| Request for citation to justify method/statement | 18 | 1.6 | 21.7 |
| Suggested literature to cite | 25 | 2.2 | 30.1 |
| **Open Science / Reproducibility** | **42** | **3.7** | **100.0** |
| Data availability: aggregated data | 3 | 0.3 | 7.1 |
| Data availability: general | 13 | 1.2 | 31.0 |
| Data availability: licence of shared data | 2 | 0.2 | 4.8 |
| Data availability: proprietary data issue | 4 | 0.4 | 9.5 |
| Data availability: raw data | 1 | 0.1 | 2.4 |
| Failed replication | 1 | 0.1 | 2.4 |
| Source code: documentation | 1 | 0.1 | 2.4 |
| Source code: general | 16 | 1.4 | 38.1 |
| Source code: reusability | 1 | 0.1 | 2.4 |
| **Overall assessment** | **118** | **10.5** | **100.0** |
| Contextualisation is (in)sufficient | 23 | 2.0 | 19.5 |
| Contribution is (not) new or useful | 33 | 2.9 | 28.0 |
| Future directions are (not) addressed | 6 | 0.5 | 5.1 |
| Research design is (in)appropriate | 29 | 2.6 | 24.6 |
| Suggested future directions | 2 | 0.2 | 1.7 |
| Topic of the paper is (ir)relevant | 25 | 2.2 | 21.2 |
| **Visualisation / Tables** | **136** | **12.0** | **100.0** |
| Add figure/table for methods description | 8 | 0.7 | 5.9 |
| Figure/Table: add additional dimension | 10 | 0.9 | 7.4 |
| Figure/Table: inconsistency in data | 6 | 0.5 | 4.4 |
| Figure/Table: not sufficiently introduced/discussed | 7 | 0.6 | 5.1 |
| Figure/Table: redundant presentation | 9 | 0.8 | 6.6 |
| Figure/Table: suggest alternate presentation | 25 | 2.2 | 18.4 |
| Figure/Table: unclear data | 23 | 2.0 | 16.9 |
| Figure/Table: unclear presentation | 48 | 4.3 | 35.3 |
| **Total** | **1,129** | **100.0** | **100.0** |

Drawing from Table 2, we see the main foci for reviewers' specific comments are:
- insufficient detail about methods (accounting for 9.6% of comments), e.g., "How do the authors treat the case when one of the m authors is affiliated to more countries than k in the ith publication? Should be stated." (Comment 14)
- a lack of clarity in the text (5.5%), e.g., "However, I have spent several hours reading the manuscript and I still struggle to understand exactly what the authors tried to do. ... the manuscript cannot be accepted for publication, unless three main forms of "unclarities" are solved ..." (Comment 107)
- requests for the justification of methodological choices (4.5%), e.g., "The authors do not explain why they choose a threshold of 25 citations." (Comment 38)
- insufficient details about the data used (4.4%), e.g., "I did not fully understand, what the dataset comprised of: WoS publications of researchers with an original affiliation with where – at European universities?" (Comment 573),
- and a lack of clarity in the presentation of visualisations (4.3%), e.g., "Figure 1: Please could you clarify in the caption that the percentiles are top 10% etc. rather than bottom 10% etc." [Comment 986]).

These results further highlight the focus of reviewers on the completeness and clarity of presentation, both in the text and visual presentation of information and results.

*3.2 Derivation of author recommendations from reviewers' comments*

The translation process of reviewer comments to recommendations to authors resulted in the 49 recommendations detailed in Table 3. Table S5 shows the broad and subcategories of comments from which the recommendations were derived. The recommendations applied to the introduction (7 items), data (8 items), methods (7 items), statistical methods (2 items), robustness (1 item), results (3 items), figures and tables (8 items), discussion (4 items), limitations (2 items), conclusions (2 items), references (2 items), and other (3 items), i.e. data availability and declarations.

Table 3. The reporting recommendations derived from peer reviews

| Section | No. | Item |
|---|---|---|
| Introduction | 1 | Authors should contextualise their study by identifying the research problem or knowledge gap that the study aims to address |
| | 2 | Authors should review and report sufficient literature so as to clearly place the study within the existing knowledge of the topic, motivate the research question(s), identify appropriate methodologies to investigate the research question, and justify the study's conclusions |
| | 3 | Authors should define acronyms at the first mention, then apply them consistently throughout the paper |
| | 4 | Authors should ensure established terms are accurately and consistently applied |
| | 5 | Authors should clearly define the concepts used in the study and apply them consistently |
| | 6 | Authors should clearly state the research question(s) that the study addresses |
| | 7 | Authors should specify the novel and relevant contribution the study makes |
| Data | 8 | Authors should clearly and thoroughly describe the dataset(s) used in the study |
| | 9 | Authors should clearly state the bibliographic databases used in the study, including which indices were included, if applicable |
| | 10 | Authors should provide the necessary information pertaining to bibliographic databases to allow readers to understand and assess the effect on the study, such as data cleaning and matching processes, descriptions of the database and its coverage and limitations, and classification structures. |
| | 11 | Authors should clearly state the dates data were collected or databases were searched, if applicable |
| | 12 | Authors should clearly state the search terms used to search bibliographic databases, if applicable |
| | 13 | Authors should clearly and thoroughly describe the bibliometric indicators used in the study, if applicable |
| | 14 | Authors should explicitly state and justify any inclusion and exclusion criteria applied to the data in their study |
| | 15 | Authors should explain the reasoning behind their selection of data and methods |
| Methods | 16 | Authors should clearly and thoroughly describe how concepts were operationalised |
| | 17 | Authors should clearly and thoroughly describe the methods used to clean the data used, including the management of outliers or unusual cases |
| | 18 | Authors should clearly and thoroughly describe the methods used to classify, disambiguate or aggregate data, such as detailing, where relevant, how author names were disambiguated, how authors were assigned to country-level affiliations, or how multiple categorisation was managed in relation to authors with multiple affiliations, or assigning journals or documents to subject classifications. |
| | 19 | Authors should clearly and thoroughly describe the methods used to calculate any bibliometric indicators, such as detailing, where relevant, at what level fractionalisation was applied, or what and how normalisations were applied |
| | 20 | Authors should clearly and thoroughly describe how any statistical tests were applied |
| | 21 | Authors should clearly and thoroughly describe the methods used to collect and analyse the data in the study |
| | 22 | Authors should, if suitable, use a flowchart or other descriptive diagram to demonstrate the methodological steps applied |
| Statistical methods | 23 | Authors should interpret statistical results, e.g., correlation coefficients or effect sizes, in accordance with accepted thresholds |
| | 24 | Authors should consider and address the assumptions and limitations of the statistical methods used, e.g., the effect of sample sizes that are very large or very small, or the appropriateness of the tests for the data available |
| Robustness | 25 | Authors should conduct and describe validity tests of both sourced data and generated results, such as checking for coverage issues, sample bias, overfitting of models, or the effects of selected thresholds |
| Results | 26 | Authors should adequately describe their results, including for example, patterns and trends in the data and validation results |
| | 27 | Authors should present sufficient detail about the results to allow readers to readily contextualise the findings, e.g., present both numbers and percentages for each result and for any corresponding overarching groups; provide median, mean or other measures of central tendency as relevant; and calculate effect sizes in addition to statistical significance where appropriate |
| | 28 | Authors should report the sizes of all samples used in the group, including relevant subgroups or overarching groups, and discuss the potential |
| | | implications of the sample size for the results in relation to the effect on any statistical tests applied and the validity and generalisability of results |
| Figures and Tables | 29 | Authors should present data in figures or tables that facilitate understanding and interpretation |
| | 30 | Authors should ensure data are consistently reported between the text, tables, and figures |
| | 31 | Authors should present figures in sufficient size and resolution for legibility |
| | 32 | Authors should ensure that figure and table titles or captions describe the figure/table and its main aspects in sufficient detail that the figure/table is able to be understood without the text |
| | 33 | Authors should ensure that all content in figures and tables, such as axis labels, titles and units, and acronyms, abbreviations, variable names, colours, symbols, or notations, is clearly explained in the figure or table |
| | 34 | Authors should clearly note when data are missing from tables or figures, such as outliers or when using subsamples |
| | 35 | Authors should clearly and consistently number figures and tables and refer to them correctly in the text |
| | 36 | Authors should ensure that all tables and figures are referenced in the text and introduced in sufficient detail |
| Discussion | 37 | Authors should discuss the study's findings in relation to the research questions and contextualise the results within the existing literature, including if applicable, any relevant theoretical frameworks |
| | 38 | Authors should ensure that the theoretical framing of the study is internally consistent with how the concepts were operationalised and the data were analysed |
| | 39 | Authors should identify and discuss potential future directions of research in the topic given the study's findings |
| | 40 | Authors should identify and discuss the theoretical/practical implications of the study's results |
| Limitations | 41 | Author should adequately discuss limitations of the data and/or methods used in the study |
| | 42 | Authors should describe any limitations introduced by the bibliographic databases used in the study, e.g., coverage of disciplines, countries, document types, etc. |
| Conclusions | 43 | Authors should draw concrete conclusions that are valid and reasonable given the scope, data, methods, and limitations of the study, and substantiated by evidence provided either by the study itself or existing literature |
| | 44 | Authors should concretely discuss the generalisability and robustness of their results |
| References | 45 | Authors should ensure that all references cited in the text are included in the reference list and that all references in the reference list are cited in the text, with complete and accurate information in both locations |
| | 46 | Authors should sufficiently reference to acknowledge ideas, methods and findings from previous work -- from the original source, if possible -- and to support the claims and statements made in the manuscript |
| Other | 47 | Authors should provide the raw data and analysis scripts underlying the study and adequate documentation of these files in a usable format with an open license for reuse by readers, where these are not constrained by proprietary licenses |
| | 48 | Authors should provide the data underlying figures in a usable format with an open license for reuse by readers |
| | 49 | Authors should declare their conflicts of interest, or state that they do not have any conflicts to report |

*3.3 Results of comparison with existing guidelines*

In comparing our list with the three guidelines, our list contained substantially more items – and thus concepts – than the guidelines. Of the 49 items in our list, 26 (53.1%) were either completely or partially present in GLOBAL, 16 (32.7%) were in PRIBA, and 18 (36.7%) were in BIBLIO, indicating that half to two-thirds of the concepts raised by reviewers were not represented in the existing guidelines. Figure 2 shows the presence of our list's items in the three guidelines, where "No" indicates absence of a comparable concept, yellow indicates partial coverage, and green indicates complete coverage and the corresponding item number(s) in the guideline. Table S6 shows the heatmap as in Figure 2 but with the item text, to facilitate comparison of the wording and content of items.

| Section | No. | GLOBAL | PRIBA | BIBLIO | Section | No. | GLOBAL | PRIBA | BIBLIO |
|---|---|---|---|---|---|---|---|---|---|
| Introduction | 1 | 2.1, 2.2 | No | No | Results | 26 | 4.1 | 11 | 15, 16 |
| Introduction | 2 | 2.1 | No | 4 | Results | 27 | No | No | No |
| Introduction | 3 | No | No | No | Results | 28 | No | No | No |
| Introduction | 4 | No | No | No | Figures & Tables | 29 | 4.3 | No | 14 |
| Introduction | 5 | 2.4 | No | No | Figures & Tables | 30 | No | No | No |
| Introduction | 6 | 2.3 | 4 | 5 | Figures & Tables | 31 | No | 12 | No |
| Introduction | 7 | No | No | No | Figures & Tables | 32 | 4.3 | No | No |
| Data | 8 | No | No | 13 | Figures & Tables | 33 | 4.3 | No | No |
| Data | 9 | 3.4 | 5a, 5b | 6 | Figures & Tables | 34 | 4.3 | No | No |
| Data | 10 | 3.4 | 5b | No | Figures & Tables | 35 | No | No | No |
| Data | 11 | 3.6 | 5c | 7 | Figures & Tables | 36 | No | No | No |
| Data | 12 | 3.5 | 7 | 7 | Discussion | 37 | 5.2 | 15 | 17, 18, 20 |
| Data | 13 | 3.11, 3.12 | 8 | No | Discussion | 38 | No | No | 18 |
| Data | 14 | 3.7 | 6 | 9 | Discussion | 39 | No | No | No |
| Data | 15 | No | No | No | Discussion | 40 | No | No | 20 |
| Methods | 16 | No | No | No | Limit. | 41 | 5.3 | 14 | 19 |
| Methods | 17 | 3.8 | No | 10 | Limit. | 42 | 3.4, 5.3 | 14 | No |
| Methods | 18 | 3.1 | No | 12 | Concl. | 43 | No | No | 18 |
| Methods | 19 | 3.11, 3.12 | 8 | No | Concl. | 44 | No | No | No |
| Methods | 20 | No | No | No | Refs | 45 | No | No | No |
| Methods | 21 | 3.9 | No | 12 | Refs | 46 | 6.3 | No | No |
| Methods | 22 | 3.7 | 10 | 13 | Other | 47 | 6.2, 6.4 | No | No |
| Methods | 23 | No | No | No | Other | 48 | No | No | No |
| Methods | 24 | No | No | No | Other | 49 | 6.1 | 17 | No |
| Methods | 25 | No | No | No | | | | | |

Figure 2. Comparison of our list's items with GLOBAL, PRIBA, and BIBLIO. "No" indicates absence of a comparable concept, yellow indicates partial coverage, and green indicates complete coverage and the corresponding item number(s).

*3.3.1 What is in our list but not in the guidelines*

There were 26 items that were unique to our list or only partially addressed in the guidelines (our items 2-4, 7, 8, 15, 16, 18, 20, 23-25, 27, 28, 30, 32-36, 39, 44-48). Many of these items constituted quite specific recommendations, such as defining acronyms (item 3), ensuring established terms (e.g., precision and recall) are used accurately and consistently (4), interpreting statistical tests in accordance with accepted thresholds (23), providing both numbers and percentages when discussing results (27), including sufficient, complete and accurate references from the most original sources (45, 46), providing the raw data, analytical scripts, and data underlying figures in a usable format with an open licence (in contrast to a data availability statement; 47, 48), and several specific details regarding the presentation of tables and figures and introducing them in the text (30, 32-36). Such details may be too fine-grained to have appeared in the literature used to derive the existing guidelines, thus excluding them. Nonetheless, their presence in our list reflects the value of reviewers in identifying mistakes – both small and large – encouraging compliance with accepted standards, and providing the perspective of the end user in terms of what is required for understanding and utilising the paper's findings.

Conversely, other items unique to our list may be considered expected standards of good academic practice and thus too broad to warrant inclusion in the bibliometrics-oriented guidelines. For instance, items such as contextualising the study in the existing literature to motivate the research questions and methods (2), specifying the novel contribution of the study (7), describing the operationalisation of concepts (16), describing any statistical tests applied (20), considering the assumptions and limitations of said tests (24), conducting and reporting validity tests (25), reporting sample sizes (28), identifying the future directions of the work (39), and discussing the generalisability and robustness of the results (44). Arguably, however, items such as reporting sample sizes should always be recommended considering that bibliometric studies very often involve immense datasets, which can influence the reliability and validity of statistical analyses.

However, we draw attention to one concept that we consider notably absent from the three guidelines: our item 18, which states "Authors should clearly and thoroughly describe the methods used to classify, disambiguate or aggregate data, such as detailing, where relevant, how author names were disambiguated, how authors were assigned to country-level affiliations, or how multiple categorisation was managed in relation to authors with multiple affiliations, or assigning journals or documents to subject classifications." We regard this item as specific and crucial to the adequate reporting of bibliometric studies. However, these topics are not explicitly addressed in the reporting guidelines, beyond recommendations to "Describe the bibliometric [data analysis] methods used" (GLOBAL items 3.1 and 3.9), and "Describe the methods used for summarizing, handling, synthesis, tabulations, or schematic displays. Describe how the data were analyzed" (BIBLIO item 14).

Similarly, as shown in Table S6, the specificity of the items in our list tends to be greater than the existing lists. For instance, regarding describing bibliometric databases, we interpreted the GLOBAL item "Describe the databases and data sources used, including any limitations" and the PRIBA item "Describe the characteristics of the data source" to capture the same concept as our item 2. However, our item provided greater context and specificity by drawing on reviewers' requested information: "Authors should provide the necessary information pertaining to bibliographic databases to allow readers to understand and assess the effect on the study, such as data cleaning and matching processes, descriptions of the database and its coverage and limitations, and classification structures." Further, reviewers were able to identify a series of specific recommendations regarding how to present figures and tables, providing more guidance than the corresponding, more general GLOBAL, BIBLIO, and PRIBA items. These examples highlight the beneficial nature of using peer review reports to derive specific details of reporting requirements.

*3.3.2 What is in the guidelines but not our list*

From the converse perspective, we also examined what items appeared in the guidelines but not in our list to identify content that our sample of reviewers did not mention. Tables S7-S9 show the mapping of our items to the GLOBAL, PRIBA, and BIBLIO lists. Of the 25 items in PRIBA, 16 (64.0%) were also included in our list and 9 (36.0%) were not. Most of the discrepancies occurred as PRIBA contained 7 items relating to the title and abstract, including identifying the study as a bibliometric analysis and specifying its coverage period in both the title and abstract, and providing the objective, main results, and their interpretation in the abstract. Notably, no items in our list pertained to the title and abstract. The remaining two PRIBA items not addressed in our list were "Specify the software package(s) used and the settings selected," and "Describe the sources of financial or non-financial support and the role of funders or sponsors."

GLOBAL contains 29 items, of which 24 (82.8%) were also identified in our list and 5 (17.2%) were not. The items from GLOBAL not covered in our list included: i) reflecting the details of the analysis in the abstract; ii) defining the units of analysis used in the study, e.g., countries, institutions, authors; iii) specifying the analytical software and parameters used; iv) "report[ing] the uncertainty/dispersion/heterogeneity depending on the type of data and analysis, and error values of bibliometric indicators;" and v) "describ[ing] the results of the bibliometric analysis techniques used". In the last case, we interpreted this item (4.2) to refer to reporting the results of applying bibliometric techniques such as normalisation or fractionalisation – rather than the results of the bibliometric analysis iteself – a concept that is not included in our list.

Of the 20 items in BIBLIO, 14 items (70.0%) were also identified in our list and 6 (30.0%) were not. As with the other guidelines, three items pertaining to the title and abstract were not present in our list. Our list also did not include BIBLIO's items pertaining to reporting and justifying the analysis' coverage period in the methods, the "[a]ssessment of papers by three authors and the use of assessing checklists," and "[s]ummariz[ing] and/or present[ing] the schematic maps and trends using an appropriate software to present citations, journals, authors, top journals, time trends, emerging literature, and any relevant indicators (as applicable)." Notably, BIBLIO item 15 provided more detailed guidance on what data to present and how than our list. We considered these details as partially covered by our item "[a]uthors should adequately describe their results, including for example, patterns and trends in the data and validation results", which we argue captures the intent behind BIBLIO's recommendations, if not the exact content.

As such, our list omits all items pertaining to the identification of characteristics of the study in the title and abstract, as well as the identification of software used for analyses, and the declaration of financial support. Also notable was that, while reviewers commonly asked authors to address their study's limitations, no mention of acknowledging its strengths was made. Consequently, our list lacks the recommendation to detail the study's strengths that is present in both GLOBAL and BIBLIO.

*3.3.3 What is common to the guidelines and our list*

Notably, a set of nine topics were common to all four lists (our items 6, 9, 11, 12, 14, 22, 26, 37, and 41). This includes recommendations that authors should i) state the study's research question(s)/objective(s), ii) state the bibliographic databases used in the analysis (and, additionally, all except BIBLIO recommend describing the characteristics also; our item 10), iii) state the dates on which data were collected, iv) state the search terms or search strategy used, v) state the inclusion/exclusion criteria applied, vi) use a flowchart to illustrate the methodological steps, vii) adequately describe the results of the study, viii) discuss the study's findings in relation to the research questions and the existing literature, ix) and discuss the limitations of the study.

An additional 14 items from our list were also present in one or two (but not all three) of the reporting guidelines (our items 1, 5, 10, 13, 17, 19, 21, 29, 31, 38, 40, 42, 43, 49). These items pertained to identifying the gap in the existing literature that the study addresses, defining the concepts used, providing thorough descriptions of the data used – particularly bibliometric indicators – and the methods used to collect, clean, and analyse data, the use of clear visualisations, discussing the limitations incurred by the use of bibliographic databases, discussing the implications of the study's findings, drawing valid and internally-consistent conclusions, and declaring conflicts of interest. Together, these 23 items could thus potentially represent the core foundation of adequate reporting of bibliometric studies.

## 4. Discussion

This study qualitatively examined 182 peer reviews of 85 bibliometric studies to identify features that reviewers praised and critiqued. Based on the reviewers' 968 comments, we quantified the focus of reviewers on particular reporting aspects, and inductively developed a set of 49 reporting recommendations for authors. We then qualitatively compared this set with three existing guidelines, GLOBAL, BIBLIO, and PRIBA, and examined the extent and focus of (dis)agreement between recommendations derived via these complementary methods.

In terms of the aspects of reporting addressed by peer reviewers, we identified 11 broad categories and 68 more specific sub-categories of themes reviewers addressed. Reviewers primarily focused on the authors' descriptions of the data and methods used in their study (29% of comments), and the general clarity of the text, the visualisations used, and the depth of the interpretation of results (~12% of comments each). As such, reviewers' comments largely pertained to critiquing and suggesting improvements to authors for the complete and clear description of ideas, data, methods, and results, both in text and visual formats.

Regarding the extent and focus of (dis)agreement between our list and the existing recommendations, we found that, while the majority of the content of the guidelines (60-80%) was addressed by the reviewers' comments, a substantial percentage of the concepts raised by reviewers were not covered in the guidelines (45-65%). The underlying cause of these discrepancies stems largely from the specific lens the reviewer takes in the publishing process, ideally positioning them to address certain features of manuscripts and precluding them from addressing others. For instance, in our sample of reviews, there were no recommendations to authors regarding either the title or abstract of their manuscript. In contrast, BIBLIO and PRIBA both specifically recommend that the study is identified in the title as a bibliometric analysis and its coverage period listed, and all three advised that the key details of the analysis are reported in the abstract. While these details may hold little relevance for reviewers and thus go unmentioned, explicitly stating this information makes such studies easier to classify in databases and discover in search queries (Page, Moher, Bossuyt, et al., 2021), motivating their inclusion in guidelines. Similarly, recommendations to declare sources of financial or other support were absent from our sample, and only one reviewer mentioned declaring conflicts of interest. This might reflect the (reasonable) expectation from reviewers that the responsibility to address such matters lies with the journal editors. As such, our recommendations are tinted by the lens of the reviewer as focused on the academic content of the manuscript. Consequently, our recommendations do not capture how guidelines may encourage reporting practices for a variety of purposes, such as ensuring discoverability and complying with journal requirements, in addition to encouraging the complete reporting of the details of the study itself.

In contrast, the recommendations developed from the reviewers' comments were more extensive than the existing guidelines and provided more detail about what should be reporting, particularly regarding the need to justify methodological choices. This reflects that the reviewers drew from specific examples of missing or insufficient information, allowing us to specify in greater detail what was recommended to be reported and the motivation behind it. Conversely, the conciseness of items in the guidelines might also reflect efforts by their developers to provide more general items to maximise the utility of the guidance across varying use cases. Further, reporting guidelines may be accompanied by Explanation and Elaboration documents that provide additional context to the recommended reporting items, reducing the need for extensive details in the guideline itself. Regardless, reviewers identified some key features of bibliometric studies that the current guidelines omit, including reporting sample sizes and specific bibliometric techniques, such as normalisation and fractionalisation. Overall, using reviewers' comments to develop recommendations resulted in a broader set of concepts authors should address than the approach currently recommended by the EQUATOR Network (n.d.). While sometimes these recommendations were perhaps too broad or specific in nature, they also identified some specific and necessary bibliometrics-oriented items that could be incorporated into future iterations of the guidelines.

*4.1 Limitations*

There are some key limitations to consider regarding our study. First, the data underlying these analyses are not systematic in that not all reviewers were asked to evaluate the manuscript on the same, specific criteria but instead commented only on the issues that occurred to them. Similarly, journal or conference peer review policies might have influenced what aspects of the submissions reviewers focused on. For instance, the STI 2023 asked reviewers to evaluate the studies' relevance for the conference, and this emerges as a high rate of positive comments regarding the studies' overall assessment in our results. However, the influence of differing policies is potentially limited as all LIS articles were published in Quantitative Science Studies, all conference submissions were to STI 2023, and only 5 studies from other journals were included, and so, essentially only two review policies are in effect. However, due to the lack of systematic review criteria, our results may not represent the true prevalence of reporting issues. Nonetheless, they are still informative about what aspects of studies reviewers focus on and to what extent.

Another consideration is that all articles and most conference submissions in our sample were accepted for publication or presentation. The focus of reviewers or the topics they raised might have been different in type or prevalence if we had been able to also include manuscripts that were ultimately rejected. Further, we only had access to the reviewer comments provided to the author, and not those provided to the editor, so we may be missing additional information from the reviews. In a similar vein, although we identified an extensive array of topics raised by reviewers, we were limited to the sample of articles that fulfilled our inclusion criteria and as such, we may not have reached saturation of the concepts that reviewers could have identified, had we included more studies.

Finally, our study is predominantly qualitative in nature and thus the categories and recommendations defined represent only our interpretation of these data. As such, we have provided all data underlying our analyses in the Supplementary Material to facilitate transparency regarding how our results and interpretations were reached and to encourage the exploration of alternatives.

## 5. Conclusions

Our study compares reporting recommendations derived from peer review comments with three existing guidelines based on published literature and expert opinion. Overarchingly, we find that the existing guidelines address up to two-thirds of the concepts identified by reviewers, but they also miss key features of bibliometric studies. Our technique of deriving recommendations from reviews thus presents a potential method for refining or supplementing guidelines in future. Notably, however, the reviewers' (appropriate) focus on academic content means this approach does not capture the dual function of reporting in ensuring, for instance, discoverability or compliance with journal policies, although it does produce a greater range and specificity of items. The recommendations derived here represent the implicit standards of the community for reporting bibliometric studies and could be used by authors in conjunction with guidelines as a detailed pre-submission checklist to ensure complete reporting of their research.

**References**
Boyack, K. W., Klavans, R., & Smith, C. (2022). *Raising the bar for bibliometric analysis.* In N. Robinson-Garcia, D. Torres-Salinas, & W. Arroyo-Machado (Eds.), 26th International Conference on Science and Technology Indicators, STI 2022. sti22143. DOI: 10.5281/zenodo.6975632.

Cabezas-Clavijo, A., Milanés-Guisado, Y., Alba-Ruiz, R., & Delgado-Vázquez, A. M. (2023). The need to develop tailored tools for improving the quality of thematic bibliometric analyses: Evidence from papers published in Sustainability and Scientometrics. *Journal of Data and Information Science, 8*(4), 10–35. DOI: 10.2478/ jdis-2023-0021.

EQUATOR Network. (n.d.). How to develop a reporting guideline. Retrieved 17 February 2025 from https://www.equator-network.org/toolkits/developing-a-reporting-guideline/.

Glänzel, W. (1996). The need for standards in bibliometric research and technology. *Scientometrics, 35*(2), 167–176. DOI: 10.1007/BF02018475.

Glänzel, W., & Schoepflin, U. (1994). Little scientometrics, big scientometrics... and beyond? *Scientometrics, 30*(2–3), 375–384. DOI: 10.1007/BF02018107.

González-Alcaide, G. (2021). Bibliometric studies outside the information science and library science field: uncontainable or uncontrollable? *Scientometrics, 126*(8), 6837–6870. DOI: 10.1007/s11192-021-04061-3.

Jonkers, K. & Derrick, G. (2012). The bibliometric bandwagon: Characteristics of bibliometric articles outside the field literature. *Journal of the American Society for Information Science and Technology, 63*(4), 829-836. DOI: 10.1002/asi.22620.

Klavans, R., & Boyack, K. W. (2022). Predicting trends in research using research community imports and exports. In: N. Robinson-Garcia, D. Torres-Salinas, & W. Arroyo-Machado (Eds.), *Proceedings of the 26th International Conference on Science and Technology Indicators (STI 2022)*. Granada: University of Granada.

Koo, M., & Lin, S-C. (2023). An analysis of reporting practices in the top 100 cited health and medicine-related bibliometric studies from 2019 to 2021 based on a proposed guidelines. *Heliyon, 9*(6), e16780. DOI: 10.1016/j.heliyon.2023.e16780.

Larivière, V. (2012). The decade of metrics? Examining the evolution of metrics within and outside LIS. *Bulletin of the American Society for Information Science and Technology, 38*(6), 12-17. DOI: 10.1002/bult.2012.1720380605.

Montazeri, A., Mohammadi, S., Hesari, P. M., Ghaemi, M., Riazi, H., & Sheikhi-Mobarakeh, Z. (2023). Preliminary guideline for reporting bibliometric reviews of the biomedical literature (BIBLIO): a minimum requirements. *Systematic Reviews, 12*. DOI: 10.1186/s13643-023-02410-2.

Moher, D., Schulz, K. F., Simera, I., & Altman, D. G. (2010). Guidance for developers of health research reporting guidelines. *PLOS Medicine, 7*(2), e1000217. DOI: 10.1371/journal.pmed.1000217.

Ng, J. Y., Haustein, S., Ebrahimzadeh, S., Chen, C., Sabe, M., Solmi, M., & Moher, D. (2023). *Guidance List for reporting bibliometric analyses (GLOBAL): A research protocol.* Open Science Framework. DOI: 10.17605/OSF.IO/MTXBF.

Ng., J. Y., Liu, H., Masood, M., Syed, N., Stephen, D., Ayala, A. P., et al. (2025). Guidance for the Reporting of Bibliometric Analyses: A Scoping Review. *Quantitative Science Studies* (online first). DOI: 10.1162/qss.a.12.

Ooms, J. (2014). *The jsonlite package: A practical and consistent mapping between JSON data and R objects.* arXiv. DOI: 10.48550/arXiv.1403.2805.

Page, M. J., Moher, D., Bossuyt, P. M., Boutron, I., Hoffmann, T. C., Mulrow, C. D., et al. (2021). PRISMA 2020 explanation and elaboration: Updated guidance and exemplars for reporting systematic reviews. BMJ, 372:n160. DOI: https://doi.org/10.1136/bmj.n160.

Wickham, H. (2022a). *rvest: Easily Harvest (Scrape) Web Pages*. (R package version 1.0.3). https://CRAN.R-project.org/package=rvest.

Wickham, H. (2022b). *stringr: Simple, Consistent Wrappers for Common String Operations*. (R package version 1.4.1). https://CRAN.R-project.org/package=stringr.

Wickham, H., Averick, M., Bryan, J., Chang, W., McGowan, L. D., François, R, et al. (2019). Welcome to the tidyverse. *Journal of Open Source Software*, 4(43), 1686. https://doi.org/10.21105/joss.01686.